\documentclass[10pt, final, journal, letterpaper,oneside, twocolumn]{IEEEtran}

\usepackage{amssymb}

\usepackage{pdfsync}

\newcounter{EquationCounter}

\usepackage{cite}

\ifCLASSINFOpdf
   \usepackage[pdftex]{graphicx}
   \graphicspath{{./}}
   \DeclareGraphicsExtensions{.pdf}
\else
\fi
\usepackage[cmex10]{amsmath}
\usepackage{amsfonts}

\begin{document}
%
\title{Spectrum Sharing Scheme Between Cellular Users and Ad-hoc Device-to-Device Users}
%
%
%


\author{\IEEEauthorblockN{Brett~Kaufman, \IEEEmembership{Student~Member,~IEEE,} Jorma~Lilleberg, \IEEEmembership{Senior~Member,~IEEE,}
\\and~Behnaam~Aazhang, \IEEEmembership{Fellow,~IEEE}}

\thanks{B. Kaufman, J. Lilleberg, and B. Aazhang are jointly with the Center for Multimedia Communication at Rice University and the Centre for Wireless Communication at the University of Oulu, Finland.  J. Lilleberg is also with Renesas in Finland.  This work is funded in part by NSF, the Academy of Finland through the Co-Op grant, and by Renesas through a research contract.}}
\maketitle

\vspace{-55pt}
\begin{abstract}
In an attempt to utilize spectrum resources more efficiently, protocols sharing licensed spectrum with unlicensed users are receiving increased attention. From the perspective of cellular networks, spectrum underutilization makes spatial reuse a feasible complement to existing standards. Interference management is a major component in designing these schemes as it is critical that licensed users maintain their expected quality of service. We develop a distributed dynamic spectrum protocol in which ad-hoc device-to-device users opportunistically access the spectrum actively in use by cellular users.

First, channel gain estimates are used to set feasible transmit powers for device-to-device users that keeps the interference they cause within the allowed interference temperature. Then network information is distributed by route discovery packets in a random access manner to help establish either a single-hop or multi-hop route between two device-to-device users. We show that network information in the discovery packet can decrease the failure rate of the route discovery and reduce the number of necessary transmissions to find a route. Using the found route, we show that two device-to-device users can communicate with a low probability of outage while only minimally affecting the cellular network, and can achieve significant power savings when communicating directly with each other instead of utilizing the cellular base station. 
%

\end{abstract}



%
\IEEEpeerreviewmaketitle

\vspace{-15pt}
\section{Introduction}
The number of wireless users is increasing at a rate faster than service providers can obtain new spectrum resources.  Coupled with the high price of purchasing new spectrum, providers need to employ new techniques in order to maximize their efficiency.  New technologies like IMT-Advanced and 3GPP Long Term Evolution (LTE) will help to satisfy the increasing demand but still more needs to be done.  Dynamic spectrum access techniques  are becoming increasingly popular as another method to meet the high demand for service.  Methods for different service providers to cooperate together to improve the overall performance of their own customers is considered in \cite{gbmidd1}.  A similar approach is taken in \cite{unused_spectrum} to redistribute excess users to frequency bands with excess capacity.  Another work is done in \cite{chen_cog_GT_resource_femto} where fixed relay stations are placed in the cell to form femtocell-hotspots. A protocol is developed in \cite{gbmiddwpmc07} where base stations take advantage of the user topology and assign resources to cellular users so that they can communicate directly with each other without the need of the base station.  

A different approach is to dynamically share licensed spectrum with unlicensed users \cite{unlic_peha,bkaufman_c3,doppler_d2d_modeselect,doppler_d2d_prospects,doppler_d2d_intjrn}.  In \cite{cellular_goldsmith}, a simple cellular model is considered to develop methods that adapt to channel conditions to reuse frequency channels among neighboring base stations.  Methods to allocate a set of frequency resources to maximize the total number of simultaneous transmissions while minimizing interference are developed in \cite{interference_andrews}.  Similarly, a pricing scheme is developed where users choose frequency channels and transmit powers to maximize their own gains while trying to reduce overall interference \cite{game_theory_resou}.  Spatial reuse techniques are used in \cite{cognitive_kamakaris} to define physical regions in a cellular network where unlicensed users could access the licensed spectrum while only causing minimal interference to the licensed users.  Different areas in literature refer to these unlicensed users as low priority, cognitive, or device-to-device users.  

We study this problem from the perspective that ad-hoc Device-to-Device (D2D) users can simultaneously operate in the same frequency spectrum as a licensed cellular radio network \cite{bkaufman_t1,bkaufman_c2,D2D_doppler_review}.  In a distributed fashion, D2D users can control their power themselves and opportunistically access the spectrum to discover routes among each other and to transmit data.  The biggest challenge in such a scheme is the interference management, specifically how to keep the level of interference the licensed macrocell users receive within the allowed interference temperature while still achieving a reliable level of performance for the unlicensed D2D users.  Much of the existing literature approaches this problem with numerous assumptions to reduce the complexity.  Interference is often managed by allocating the unlicensed users with frequency resources that are disjoint to those of the licensed network, either through a completely different frequency band or a subset of the same band that is currently not in use \cite{cellular_adhoc}.  Alternatively, the locations of interfering licensed users are assumed to be known and can be used to exploit spatial holes in a network's frequency resources \cite{location_known}.  Methods like \cite{decentralized_control}, which assume a centralized controller with full knowledge of the network, are even more common.  


In our approach, we develop a distributed dynamic spectrum protocol to enable device-to-device communication.  We study the problem on a coarse time scale where the topology and channel conditions are fixed for an entire frame.  Furthermore, we focus on the power levels of signals allowing us to abstract away the specific modulation and coding schemes used.  We intend for D2D users to utilize statistical estimates of the channel gains to set a transmit power level that will be within the allowed interference temperature of the cellular network.  Then using the calculated transmit power, two D2D users will attempt to discover either a single-hop or multi-hop route connecting each other.  We utilize network information in the discovery packet to improve both the success and efficiency of the route discovery.  Random access techniques are used to ensure that only one D2D user accesses the spectrum at a given time.  Once a route is found between the two users, we are able to quantify the D2D link quality in terms of the probability of outage and power savings.  

Our results show that the network information in the discovery packets decreases the probability of failure in finding a route while significantly reducing the number of transmissions necessary to discover a route.  With the framework in place to find routes between D2D users, we then derive the probability of outage for a link between any two D2D users.  We consider perfect channel inversion in the power control to calculate an analytical lower bound on the outage probability and show that our distributed power control using statistical estimates performs well compared to the lower bound.  We then give the outage probability for the cellular users and show that a large improvement in D2D performance comes at the cost of only a small loss in cellular performance.  Furthermore, D2D users achieve significant power savings by communicating over shorter low power routes instead of utilizing the potentially distant base station.

The remainder of this paper is organized as follows.  In the next section we define the cellular model that we consider.  Section \ref{sec:discovery} describes in detail how the power control and route discovery in the protocol work and give simulation results showing the performance of the route discovery.  In Section \ref{sec:singlehopd2d} we present the analytical expressions for the probability of outage and give results that quantify the performance of our device-to-device scheme.  Concluding remarks and future extensions appear in Section \ref{sec:conclusion}.  
\section{Network Architecture}
\label{networkmodel}
\subsection{Infrastructure and User Model}
The network considered consists of seven circular cells of radius $R$ with a base station (BS) equipped with an omni-directional antenna located at the center of each cell.  We focus on the uplink frame of the system and assume it to be divided into $N_C$ orthogonal channels.\footnote{The technique used to orthogonalize the $N_C$  channels can vary for different systems, i.e. separation in frequency, time, or code (OFDM, TDMA, or CDMA). Our work is feasible for any separation technique as long as each of the $N_C$ channels only serves one cellular user.  } We consider the same $N_C$ channels are available for use in each cell.  For a cellular link to be established with the base station, a minimum SINR of $\beta_B$ is required.  We assume that there exists a margin $\kappa$ in the required SINR at each base station.  This margin corresponds to the allowed interference temperature of the network and is a common design feature of wireless systems as there is always some fluctuation in the overall interference temperature of the network \cite{Martin-Sacristan2009On-the-Way-Towa}.  


The first type of user is a macro user (MU) and communicates by establishing a link with the nearest base station and having their information relayed to their intended destination.  Macro users access the base station using standard control signaling found in today's cellular systems.  There are $N_M$ active macro users uniformly distributed in each cell and we assume that there is only one active macro user per channel.  This gives the relation $N_M = N_C$ which is in place to ensure that all of the $N_C$ channels are actively in use in each cell and that there are no spare channels to be reallocated elsewhere.  This assumption holds for all cells in the network, thus neighboring cells have no channels to lend each other for cell-edge users.  

The second type of user, a D2D user, communicates directly with each other in a distributed  ad-hoc fashion over one or more hops without any assistance by the base station.  All D2D users are uniformly distributed within a single randomly located circular cluster of radius $r$ where we assume $r \ll R$.  This user cluster is distributed such that the entire area of the cluster is within the boundary of the macrocell.  Furthermore, we choose two D2Ds in the cluster at random where one is a D2D source (S) with information intended for the other, a D2D destination (D).  If a single-hop link between the source and destination is not available, we assume that there are $N_D$ idle D2D users willing to serve as relays in a multi-hop route.  We intend for this type of user and topology to be representative of those that you would find in a school campus, hospital, or commercial center where two communicating users are often located near each other and there is often a high density of idle users. 

D2D users communicate with each other on the same frequency channels used by macro users, however their use of those channels cannot cause the SINR of an active cellular link to fall by more than the allowed $\kappa$.  To meet this requirement, we assume that D2D users know the value of $\kappa$.  Base stations in current cellular systems periodically broadcast information to users and the value of $\kappa$ could be included in that standard control signaling.  D2Ds utilize CSMA/CA to randomly access the channels and will discover each other using a protocol described in Section \ref{sec:discover_protocol}.  Finally, for a D2D link to exist, a minimum SINR of $\beta_{D}$ must be achieved between a transmitting D2D and a receiving D2D.  

As a final remark, we note that D2D users and macro users only differ in their modes of communicating with each other, either direct or though the base station.  In fact both classes of users would be composed of the same type of wireless devices.  D2D users are simply macro users who could not be served by the base station.

\subsection{Channel Model}

We present our channel model in the context of the network defined above.  We consider three arbitrary users: a transmitter $i$, a receiver $j$, and an interferer $k$.  We assume a pathloss dominated channel with multiplicative fading and additive white Gaussian noise.  The large-scale fading is determined by the Euclidian distance $d_{ij}$ between two users $i$ and $j$ and the pathloss exponent $\alpha$.  A Rayleigh random variable $f_{ij}$ determines the small-scale fading between the same two users. We are primarily interested in the power of user's signals and the corresponding SINR of their links and thus define user $j$'s SINR as
\begin{equation}
\Gamma_j = \dfrac{P_{T_{i}}d_{ij}^{-\alpha}h_{ij}}{\sum\limits_k P_{T_{k}}d_{kj}^{-\alpha}h_{kj} + \sigma^2},
\end{equation}
where $P_{T_{i}}$ is the power used by the transmitter, $d_{ij}^{-\alpha}$ is the pathloss for the link between the transmitter and receiver, and $h_{ij}=|f_{ij}|^2$ is the channel gain.  Similarly, $P_{T_{k}}$ is the power used by the $k$'th interferer and $d_{kj}^{-\alpha}$ and $h_{kj}=|f_{kj}|^2$ are the pathloss and channel gain for the link between the $k$'th interferer and the receiver.  We assume that all users observe the same noise power of $\sigma^2$.  We will use the subscripts $M$, $S$, and $D$ to denote the different parameters for the macro user, source, and destination.  Finally, we assume all users know the pathloss of their respective links with the base station.  The knowledge of the large-scale fading will help users to manage their interference through a power control scheme and the fast fading will be used to analyze the performance of that scheme under random channel conditions.  
\section{Device-to-Device Communication}
\label{sec:discovery}
We propose a dynamic spectrum access protocol in which Device-to-Device users can communicate directly with each other using the same frequency resources as a simultaneously active uplink between a macro user and the base station.  This protocol is opportunistic as a link between two D2D users can only be utilized if their use of the spectrum  stays within the interference temperature of the network and does not cause the SINR of the cellular link to decrease by more than the allowed margin.  To best accomplish this, we only allow D2D users to communicate with each other during the uplink frame of the network.  During the uplink transmission phase, only the stationary base station will receive interference from the D2Ds.  Macro users will not receive any interference from the D2Ds as they will be uplinking to the base station.  If D2D users communicated during the downlink, interference would be seen at every macro user in the system.  It is impractical to assume that the macro user locations and channel conditions are known by each D2D user making it extremely difficult for D2D users to control the interference they cause.  We note that the same practical reasons prevent D2Ds from accurately controlling their power to adjust for interference from active macro users.  Thus, macro user interference will not affect the power control but will be considered in D2D link quality when the SINR thresholds are evaluated.  

There are two main steps in our protocol.  First, is the power control for D2D users.  Because D2D's use of the macro user's bandwidth is constrained by how much interference they cause, the power control will be the main determining factor in the protocol's performance.  Once an allowed transmit power has been calculated, the second step will be for a D2D user to discover either a single-hop or multi-hop route to their intended destination.  In this section, we will outline the steps for each of these components.  

\subsection{Power Control for Device-to-Device Users}
\label{sec:power_control}

Recall that a minimum SINR of $\beta_B$ is required for a macro user link to exist with the base station and that there can be at most a $\kappa$ change in the base station's SINR due to interference from a transmitting D2D user.  If D2D users control their interference perfectly, a macro user with perfect power control will achieve the required SINR of $\beta_B$.  This result is obtained by the macro user scaling its own transmit power by $\kappa$.  Thus in the absence of the D2D interference, a macro user link will achieve an SNR, or SINR with zero interference, of $\kappa\beta_B$.  We can see the effects of $\kappa$ in the macro user link by looking at the SNR at the base station, where after rearranging terms, gives a bound on the transmit power of macro users as
\begin{eqnarray}
\dfrac{P_{T_{M}}d_{MB}^{-\alpha}h_{MB}}{\sigma^2} &\geq& \kappa \beta_{B} \nonumber
\\
P_{T_{M}} &\geq& \kappa  \beta_{B} d_{MB}^{\alpha}h_{MB}^{-1} \sigma^2 \triangleq P_{T_M}^{min},
\label{eq:ptc_bound}
\end{eqnarray}
where $d_{MB}^{-\alpha}$ and $h_{MB}$ are the pathloss and channel gain between a macro user and the base station.  Assuming prefect knowledge of the channel gain, the bound in (\ref{eq:ptc_bound}) gives a transmit power for macro users such that the probability of outage will be zero.  In this work, we will focus on the performance of the D2D users, so we assume that macro users are power controlled by the base station with an error free estimate of the channel gain.  In practice, the estimate may contain some error causing the macro user to go into outage.  However, that outage will be independent of the D2D's use of the channel.  
%

We now evaluate the SINR of a macro user link that is interfered by a single random D2D user. Without loss of generality, we refer to that D2D as the source $S$.  If we take $P_{T_{M}}$ to be the minimum allowed in (\ref{eq:ptc_bound}), after rearranging terms, we get a bound on the transmit power of a D2D user as
\begin{eqnarray}
\dfrac{P_{T_{M}}d_{MB}^{-\alpha}h_{MB}}{P_{T_{S}}d_{SB}^{-\alpha}h_{SB}+\sigma^2} &\geq& \beta_{B} \nonumber
\\
P_{T_{S}}^{max} \triangleq(\kappa-1)d_{SB}^{\alpha}\sigma^2h_{SB}^{-1} &\geq& P_{T_{S}},
\label{eq:ptd_bound}
\end{eqnarray}
where $d_{SB}^{-\alpha}$ and $h_{SB}$ are the pathloss and channel gain between the source and the base station.  Assuming a D2D has perfect knowledge of $\kappa$ and $h_{SB}$, the bound in (\ref{eq:ptd_bound}) gives a transmit power that a D2D can use and not cause a macro user to go into outage.  We assume no coordination between D2Ds and the base station so $h_{SB}$ will be unknown and must be estimated.  We use a statistical estimate and assume $h_{SB}$ is estimated by the mean of the fading, and define $\widehat{h_{SB}} \triangleq \mathbb{E}[h_{SB}]$.  Using that estimate, a D2D can calculate a usable maximum transmit power as $P_{T_{S}}^{max^*} \triangleq (\kappa-1)d_{SB}^{\alpha}\sigma^2(\widehat{h_{SB}})^{-1}$.  

The accuracy of $\widehat{h_{SB}}$ in estimating $h_{SB}$ will determine how much interference the base station receives from a transmitting D2D user.  When $\widehat{h_{SB}} < h_{SB}$, the calculated maximum $P_{T_{S}}^{max^*}$ will exceed the true maximum $P_{T_{S}}^{max}$ and the macro user will go into outage.  Conversely, when $\widehat{h_{SB}} > h_{SB}$, $P_{T_{S}}^{max^*}$ will be lower than the true maximum.  This in turn means that a D2D could potentially use $P_{T_{S}}^{max} - P_{T_{S}}^{max^*}$ additional transmit power and still not cause a macro user to go into outage.  This can be exploited to allow for additional scaling of the D2D link power to improve D2D link quality at no cost to the macro user link quality.  Thus far, $P_{T_{S}}^{max^*}$ has been contolled to minimize the interference they cause to the base station.  Ideally, the power control for a D2D user should also consider the link used to reach the intended destination $D$, specifically the fading $h_{SD}$.  We know that for the source transmitting with power $P_{T_{S}}^{max^*}$, a correctly received packet at the destination will have power $P_{R_{D}}=P_{T_{S}}^{max^*} d_{SD}^{-\alpha} h_{SD}$.  The additional power control mentioned just above could be realized in the form of an estimate of the fading, denoted as $\widehat{h_{SD}}$.  

Using that estimate, channel inversion can be used in the power control of the D2D link.  The calculated maximum power of $P_{T_{S}}^{max^*}$ can be scaled to set a new usable transmit power for D2D users as $P_{T_{S}}^{*} = P_{T_{S}}^{max^*} (\widehat{h_{SD}})^{-1}$.  Numerous works in current literature show that channel estimation is feasible but the distributed nature of the D2D communication may prohibit any additional overhead for channel estimation.  As such, we use a constant estimate and assume $\widehat{h_{SD}} \triangleq \mathbb{E}[h_{SD}]$.  This is a practical choice for the estimate and is equivalent to just power controlling over the pathloss and ignoring the fast fading effects.  

For analysis purposes, we will consider two other cases of $\widehat{h_{SD}}$  that are impractical for a real protocol but are important  in order to bound the performance of the system.  It can be easily shown that $P_{T_{S}}^{*} \leq P_{T_{S}}^{max^*}$ is achieved with $\widehat{h_{SD}} \geq 1$.  Thus setting $\widehat{h_{SD}} = \max(1, h_{SD})$ will result in a $P_{T_{S}}^{*}$ that does cause macro user outage due to the additional power control added for the D2D link.  We note that this choice of estimate is equivalent to truncated channel inversion with a truncation threshold of 1 \cite{goldsmith_book}.  It was shown above that in some cases $P_{T_{S}}^{max^*}$ will be less than $P_{T_{S}}^{max}$ and additional power could be used.  In this scenario, it is not necessary for $P_{T_{S}}^{*} \leq P_{T_{S}}^{max^*}$ and in fact some $\widehat{h_{SD}} < 1$ could be used.  No exact threshold can be solved for when $\widehat{h_{SD}} < 1$ is feasible as it will depend on a particular realization of $h_{SB}$ and  $P_{T_{S}}^{max}$. Based on this, we consider perfect channel inversion, $\widehat{h_{SD}} = h_{SD}$, which analyzes the system as if there were no interference constraints or limits on transmit power levels. 

After the power control has been completed, the source is able to set its transmit power to 
\begin{equation}
P_{T_{S}}^{*} = (\kappa-1)d_{SB}^{\alpha}\sigma^2(\widehat{h_{SB}})^{-1}(\widehat{h_{SD}})^{-1}.
\label{eq:d2dusedpower}
\end{equation}  
The power in (\ref{eq:d2dusedpower}) assumes a single-hop link between the D2D source and destination.  When a multi-hop route is formed, the same power is used by changing the subscripts $S = i$ and $D = j$ to denote a link between the $i$'th and $j$'th D2D users.  

Using the mean of the fading as the two estimates gives a practical method for D2Ds to set their own transmit power with minimal overhead.  In Section \ref{sec:singlehopd2d}, the performance of the system will be analyzed in terms of the probability of outage and will be bounded by the two analytical scenarios of perfect channel inversion and truncated channel inversion.  Next we will describe the second step of the protocol where D2Ds use a practical discovery method to establish routes using the transmit power just derived.

\subsection{Distributed Route Discovery for Two-way Device-to-Device Communication}  
\label{sec:discover_protocol}
In our work, we use the Dynamic Source Routing (DSR) protocol\cite{dsr_original} to discover D2D links in our network.  DSR is a source initiated packet based discovery protocol.  The DSR protocol floods the network with discovery packets and in doing so, exchanges the address of relay nodes in the network so the destination will have a virtual map of how to reach the source.  Results in \cite{flood_routing_loh,flood_jun} show that DSR, as well as other flooding techniques, are beneficial in the sense that there is diversity in the discovery message by traversing more than one link.  The same results show that by flooding route information through the network, both nodes in the route and near to the route, learn of the route's existence.  If node mobility or adverse channel conditions cause the route to break, nearby nodes can easily help repair the route.  However, flooding can have negative effects especially in terms of overhead and interference.  Work in \cite{dsr3} shows that the flooding rules in DSR help to keep the routing overhead small.  Specifically, discovery packets are only allowed to traverse small areas of the network and will never traverse the same link twice, preventing loop problems.  Furthermore, D2Ds employ CSMA/CA to ensure that only one D2D accesses the channel at a given time.  By enforcing these rules, the number of discovery transmissions decreases therefore reducing both the overhead and the interference effects of DSR.  

In our implementation of the DSR protocol, we intend for forwarding relays to add their transmission power and measured interference power to the discovery packet in addition to their address.  To begin, the D2D source broadcasts a packet intended for the destination and includes in the packet its transmission power $P_{T_{S}}^*$, as calculated from (\ref{eq:d2dusedpower}), and its own measured interference power, $I_{S}$.  Without loss of generality, assume the packet is received by the $j$'th D2D, denoted as node $j$.  To see the effects of including those two powers in the discovery packet, we look at
\begin{equation}
\Gamma_{j} = \dfrac{P_{R_{j}}}{I_{j} + \sigma^2},
\label{eq:measuredSINR}
\end{equation}
where $\Gamma_{j}$ and $P_{R_{j}}$ are the measured SINR and received power at node $j$.  $I_j$ represents the measured sum interference power which accounts for both in-cell and out-of-cell interferers at node $j$.  Using (\ref{eq:measuredSINR}) and the fact that $P_{R_{j}} = P_{T_{S}}^{*} d_{Sj}^{-\alpha} h_{Sj}$ we can write
\begin{equation}
d_{Sj}^{-\alpha} h_{Sj}= \dfrac{\Gamma_{j}(I_{j}+\sigma^2)}{P_{T_{S}}^{*}}, 
\label{eq:channel_power}
\end{equation}
which solves for the combined pathloss and channel gain.  Node $j$ knows $P_{T_{S}}^{*}$ from the discovery packet, and $\Gamma_{j}$, $I_{j}$, and $\sigma^2$ are measured values, so node $j$ can calculate the combined pathloss and channel gain.  In a similar fashion, the required SINR constraint of $\beta_D$ for node $j$, with transmit power $P_{T_{j}}^{*}$ to communicate back to the source, can be rewritten to show
\begin{eqnarray} 
\dfrac{P_{T_{j}}^{*} d_{jS}^{-\alpha} h_{jS}}{I_{S}+\sigma^2} &\geq& \beta_D \nonumber
\\
P_{T_{j}}^{*} &\geq& d_{jS}^{\alpha}h_{jS}^{-1}(I_{S} + \sigma^2)\beta_{D} \triangleq P_{T_{j}}^{min},  
\label{eq:ptdbound}
\end{eqnarray}
which gives the minimum transmit power that node $j$ must use to communicate with the source.  We assume symmetric channels such that $d_{Sj}^{-\alpha} h_{Sj} = d_{jS}^{-\alpha} h_{jS}$ and the discovery packet contains the interference power seen by the source, thus node $j$ can calculate $P_{T_{j}}^{min}$.  If $P_{T_{j}}^{*} \geq P_{T_{j}}^{min}$ is satisfied, node $j$ knows that its packet will be received by the source even before it is transmitted.  In certain topologies and channel conditions, there will be no value for $P_{T_{j}}^{*}$ in which a two-way D2D link can be sustained and the link would be considered in outage.  

If node $j$ is the D2D destination, then a single-hop route exists with the D2D source.  If a single-hop route does not exist, then node $j$ can continue the discovery process and serve as a relay.  However, it will only continue the discovery if it knows that a two-way link exists with the D2D source.  It re-broadcasts the discovery packet adding its own transmission power and interference power.  Any node receiving it can repeat the steps in (\ref{eq:measuredSINR}), (\ref{eq:channel_power}), and (\ref{eq:ptdbound}) to determine if a two-way link exists.  Once the D2D destination receives the packet, it will have a list of relay nodes that form a multi-hop route with the D2D source.  

\begin{figure}[htp]
\center
 \includegraphics[scale=.8]{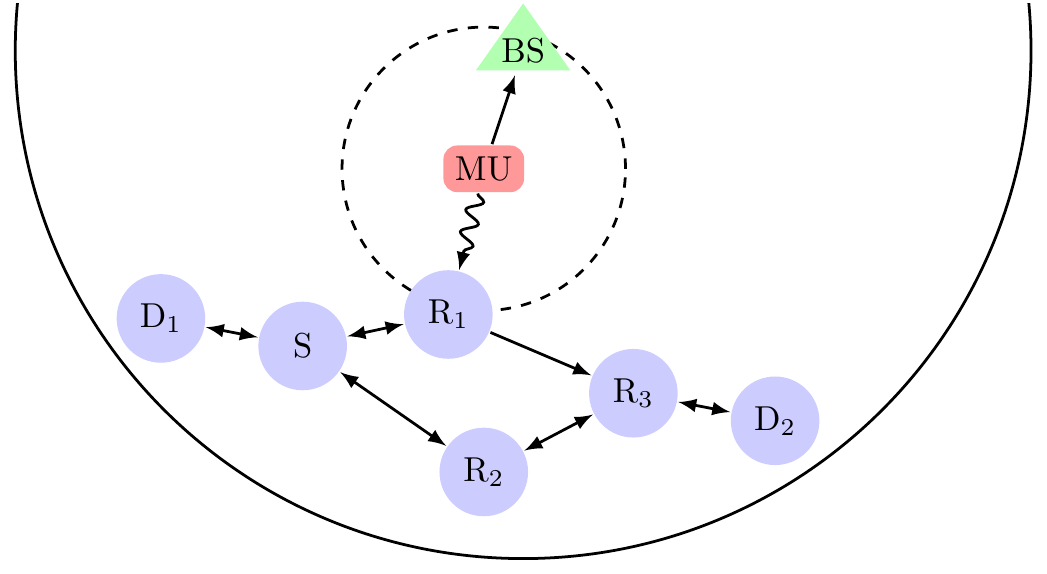}
 \caption[]{An example realization of what a cellular network with an underlaid D2D network may look like.  The source (S) communicates over a single-hop if possible, as to D$_1$, or uses idle D2D users R$_{i}$ as relays over a multi-hop route, as to D$_2$.  The interference from the macro user (MU) causes too much interference for a two-way route to be used with relay R$_1$. } 
  \label{fig:example}
\end{figure}

We now outline how a multi-hop route connecting two random D2Ds can be discovered and refer to Fig.~\ref{fig:example} as reference.  Assume that the source S wants to communicate with the destination D$_1$ by using the same channel as the active macro user MU.
The source transmits a discovery packet intended for D$_1$ at a power level of $P_{T_{S}}^{*}$.  D$_1$ is sufficiently far away from the interfering macro user to receive the packet and uses the values of $P_{T_{S}}^{*}$ and $I_S$ contained in the packet to determine that a two-way single-hop link exists with S.  Now assume that S wants to communicate with D$_2$.  Each relay R$_i$ would forward the discovery packet intended for D$_2$ after adding their own transmit and interference power to the packet.  A two-way route between R$_1$ and R$_3$ could not be used due to strong interference from the macro user.  Using the powers in the discovery packet from R$_1$, R$_3$ would be able to determine that a two-way link does not exist with R$_1$ and would not forward the discovery packet.  This reduces the number of transmissions necessary for discovery and increases the chances of discovery packets along two-way routes reaching the destination.  A two-way route could be established using R$_2$ instead.  

In both the single-hop and multi-hop routes discussed above, the destinations need to be able to reply back to the source with the correct routing.  By doing so, the source knows a route exists to its intended destination and its message should be successfully delivered.  By including the transmission and interference powers in the discovery packet, D2D users get important network information about links with their neighbors.  The network information will improve the likelihood that a two-way route is discovered and lower the number of transmissions necessary to do so.  

\subsection{Route Discovery Simulation Results}
\begin{figure}[htp]
\center
  \includegraphics[scale=0.8]{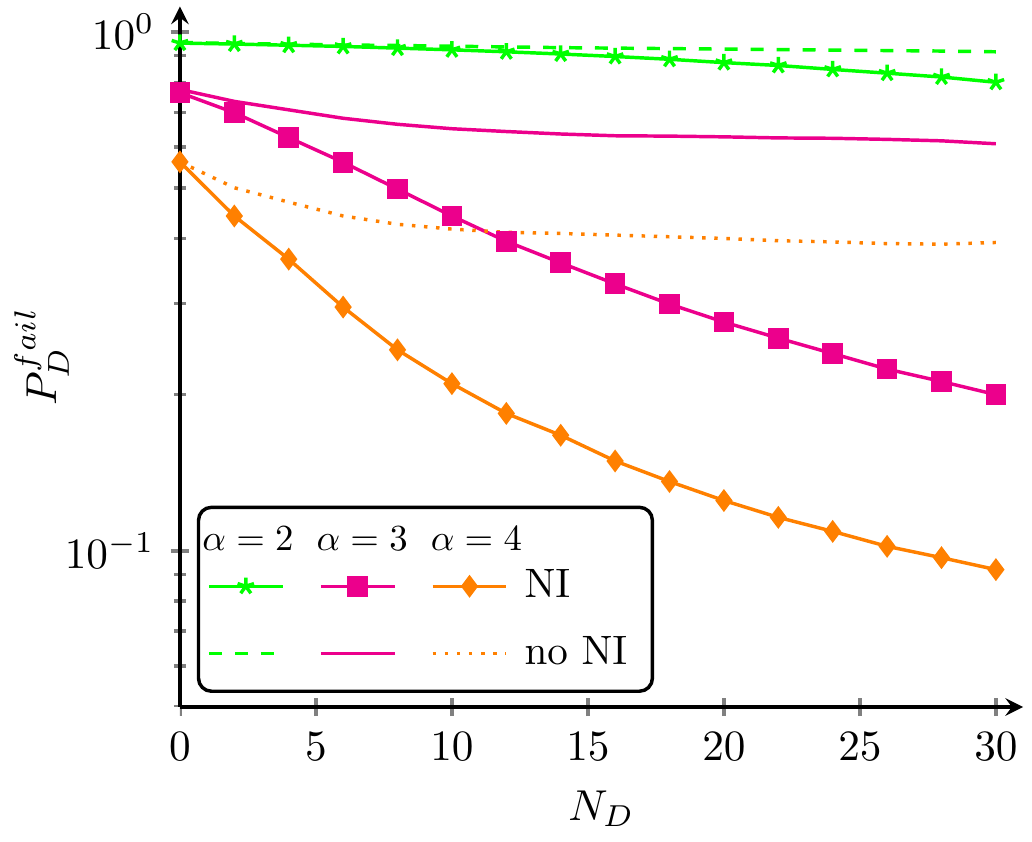}
  \caption[]{The probability of failure in discovering a two-way D2D route ($P^{fail}_{D}$).  The effects of network information (NI) in the discovery packet are considered.  } 
  \label{fig:prob2way}
\end{figure}

Our network model was simulated for $10^6$ random topologies in MATLAB. The various network parameters used for the simulations are shown in Table~\ref{table:vars}.  
\begin{table}[htp]
\begin{center}  
\caption{Network Parameters} 
\centering 
\begin{tabular}{|c||c|} 
\hline   
\textbf{System Parameters} & \textbf{Value} \\   
\hline\hline 
Cell Radius ($R$) & 2000m\\
\hline
Cluster Radius ($r$) & 500m \\
\hline
Number of Channels ($N_C$)& 30 \\ 
\hline
Number of MUs ($N_M$)& 30 \\ 
\hline
Noise ($\sigma^2$)& -104 dBm  \\ 
\hline
Minimum BS SINR ($\beta_{B}$)& 10 dB  \\
\hline  
Minimum D2D SINR  ($\beta_D$)& 5 dB  \\ 
\hline
Interference Margin at BS ($\kappa$)& 3 dB  \\
\hline
Mean of Rayleigh Fading ($E[h_{ij}]$)& 1\\
\hline
\end{tabular} 
\label{table:vars}  
\end{center}
\end{table}
We note that these simulation results show the performance of the center macrocell only but take into account the interference effects from the six other surrounding macrocells.  We show in Fig.~\ref{fig:prob2way} the probability of failure in discovering a two-way D2D route, denoted as $P^{fail}_{D}$, versus the number of available D2D relays, $N_D$, in the cluster.  We note that for a given value of $N_D$, not all relays participate in the route connecting the source and destination.
We immediately see that as $N_D$ increases, the failure rate decreases.  Relays willing to forward information for the D2D source can help overcome high attenuation channels due to large distances and random fading.  We also note that as $\alpha$ increases, $P^{fail}_{D}$ decreases.  Even though each hop in the D2D route sees higher attenuation, interference from cellular users is lower, and D2Ds interfere with the BS less allowing them to transmit at a higher power.  Finally, the figure shows that using the network information in the discovery packet can significantly improve the probability of discovering a two-way route.  

\begin{figure}[htp]
\center
  \includegraphics[scale=0.83]{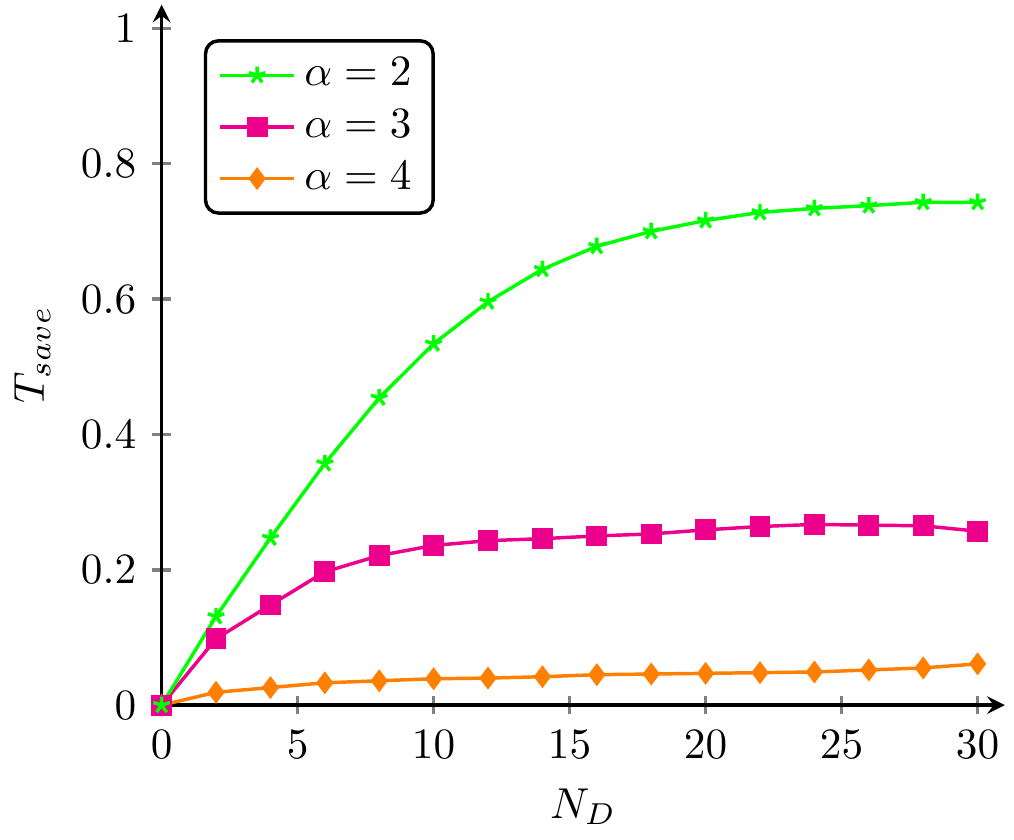}
  \caption[]{The percentage savings in the number of discovery transmissions made ($T_{save}$) when network information is used in the discovery packet.} 
  \label{fig:trancount} 
\end{figure}

Recall that the network information (NI) reduces the number of discovery transmissions, and thus the overhead, needed to establish a route.  To quantify the savings in discovery transmissions 
\begin{equation}
T_{save} = \dfrac{T-T_{NI}}{T},
\end{equation} 
we compare the average number of transmissions used to discover a route when the NI is included in the discovery packet, denoted by $T_{NI}$, to the average number of transmissions used when the NI is not included in the discovery packet, denoted by $T$.  Fig.~\ref{fig:trancount} plots $T_{save}$ and shows increasing savings as $\alpha$ decreases.
For small $\alpha$, there is more interference in the network which increases the likelihood  that D2D links will be one-way.  This in turn results in more D2D relays not forwarding their received discovery packets.  The gains are evident most at larger $N_D$ when there could be potentially many relays forwarding discovery packets received over one-way links.  

\section{Outage Analysis of Device-to-Device Communication}
\label{sec:singlehopd2d}
With a practical framework in place for D2D users to first power control themselves and then to establish routes among themselves, we now derive the probability of outage for a link between two randomly placed D2D users.  Our approach is geometric in nature and is motivated by the various random distances in the model considered.  We derive the D2D outage probability using both the distance and fading channel probability distributions which allows us to consider all D2D locations in the macrocell and all possible channel conditions between users.  In order to derive the single-hop analytical expression in Section~\ref{sec:single_hop_outage}, we assume a single macrocell system where the interference from surrounding cells is ignored.  Incorporating the additional random distances from neighboring macrocells in the analysis makes a solution intractable.  However, those interference effects are fully accounted for in the simulation results of Section~\ref{sec:multihop_simulation}.  We will show that when a multi-hop route is used to connect the D2D source-destination pair, the average number of hops is low and large power savings can be realized.  

\subsection{Single-hop Probability of Outage Derivation}
\label{sec:single_hop_outage}
\begin{figure*}[!t]
\normalsize
\setcounter{EquationCounter}{\value{equation}}
\setcounter{equation}{11}
\begin{equation}
\label{eq:outage_general}
P^{out}_{D|C_{i}} = 1- Pr[d_{SD} \leq d_{max}] = 1 - \int\limits_{d_{ij},h_{ij}}  \dfrac{A_{INT}}{\pi r^2} \; p(h_{MB},h_{MD},h_{SD},d_{SC},d_{SB},d_{MD},d_{MB}),
\end{equation}
\setcounter{equation}{\value{EquationCounter}}
\hrulefill
\vspace*{-20pt} 
\end{figure*}
\begin{figure}[htp]
\center
 \includegraphics[scale=.54]{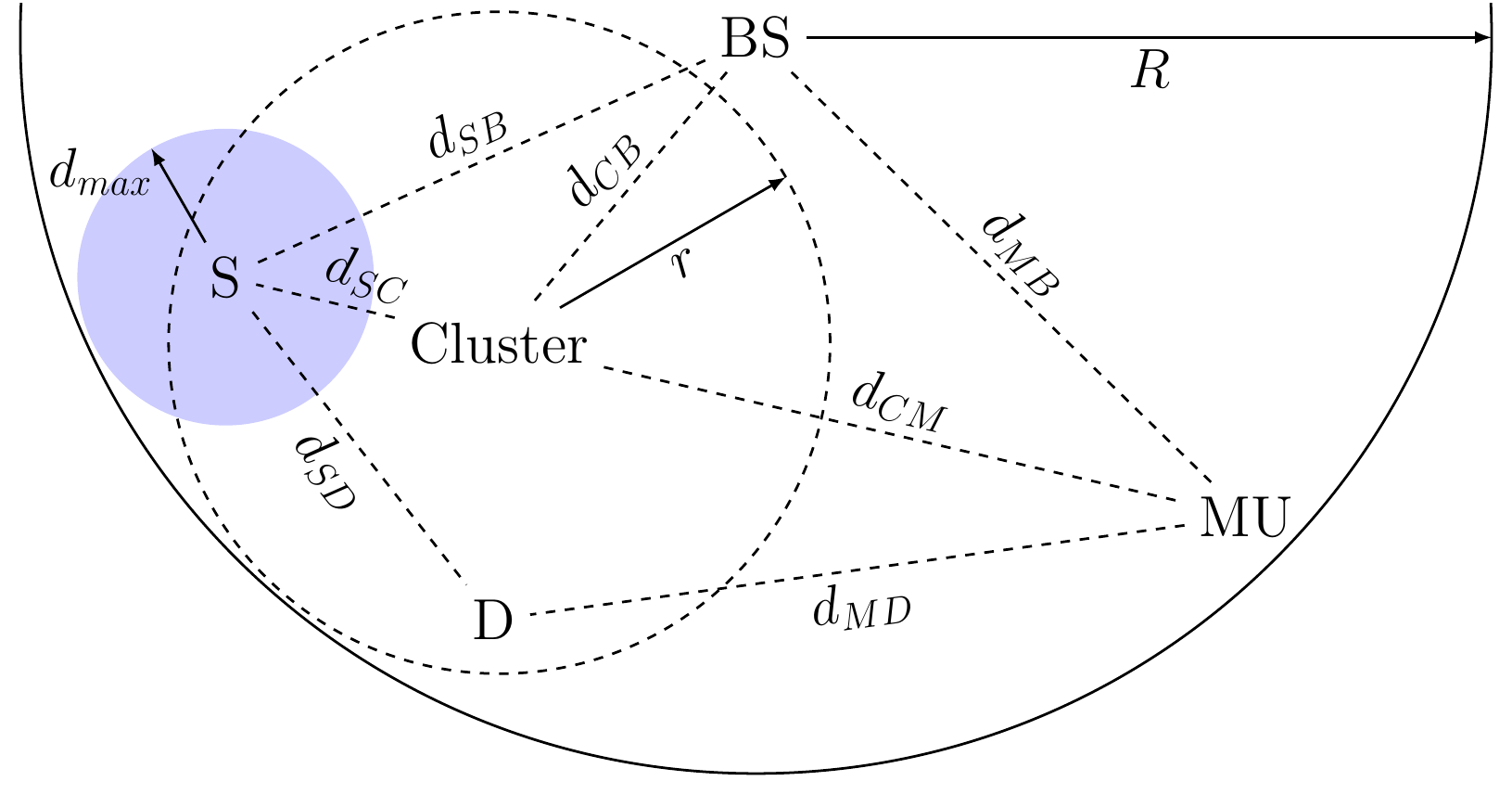}
  \caption[]{Network topology realization presenting the various random distances in the model.} 
  \label{fig:analyticalmodel}
\end{figure}

We refer to Fig.~\ref{fig:analyticalmodel} as an example network topology realization as we derive the outage probability of a single-hop link between two randomly placed D2Ds, a probability defined as $P^{out}_{D}$.  As a first step, we will derive the outage probability on all channels in terms of the outage probability for a single channel.  Recall that the available bandwidth is divided into $N_{C}$ orthogonal channels and D2Ds are able to access any of them.  The orthogonality of those channels results in a link's outage on channel $C_{i}$ being independent of and identically distributed to a link's outage on channel $C_{j}$ for $i\neq j$.  Thus we can easily write 
\begin{equation} 
P^{out}_{D} = \Big(P^{out}_{D|C_{i}}\Big)^{N_C},
\label{eq:m2a_multi_channel_prob3}
\end{equation}
which gives the probability of outage for a single-hop D2D link on $N_C$ orthogonal channels.  

Assuming a given channel $C_i$, the second step is to derive the probability of outage $P^{out}_{D|C_{i}}$.  We will use the geometry of the model to define the outage in terms of the random distance between the source and destination.  Recall we consider a single D2D source-destination pair separated by a distance $d_{SD}$.  We know for a D2D link to exist, the SINR at the destination must be above the required threshold $\beta_{D}$.  If we look at the SINR equation for the D2D link and set the macro user and D2D user transmit powers as $P_{T_M} = P_{T_M}^{min}$ and $P_{T_S}=P_{T_S}^*$, as derived in Section~\ref{sec:power_control}, after rearranging terms, we obtain 
\begin{eqnarray} 
\beta_{D} \!\!\!\!&\leq& \!\!\!\dfrac{P_{T_{S}}d_{SD}^{-\alpha}h_{SD}}{P_{T_{M}}d_{MD}^{-\alpha}h_{MD}+\sigma^2} \nonumber
\\
d_{SD} \!\!\!\!&\leq&\!\!\!\!\! \Bigg( \dfrac{d_{MD}^{\alpha}(\kappa -1)d_{SB}^{\alpha}h_{SD}h_{MB}(\widehat{h_{SB}})^{-1}}{\beta_D\widehat{h_{SD}}\big[\kappa \beta_{B} d_{MB}^{\alpha}h_{MD} + d_{MD}^{\alpha}h_{MB}\big]}   \Bigg)^{\frac{1}{\alpha}} \!\triangleq d_{max}, \nonumber
\\
\label{eq:d}
\end{eqnarray}
which gives an upper bound on the allowed distance between the D2D source and destination as a function of the network parameters in the model.  The distance $d_{max}$ is the maximum transmission distance of the source and defines a region around the source, shown by the shaded area in Fig.~\ref{fig:analyticalmodel}, in which the destination must be located in order to satisfy the required SINR $\beta_D$.\footnote{In Fig.~\ref{fig:analyticalmodel}, the shaded region is a circle corresponding to a distance based pathloss system.  In our combined pathless fading channel model, the region will be approximately circular.}  Thus, the probability of a link satisfying the SINR requirement and not being in outage, $Pr[d_{SD} \leq d_{max}]$, is the ratio of all the feasible locations of the destination that result in a successful link, which is the coverage region of the source, to all possible destination locations, the area of the entire cluster.  Using a result from \cite{geometry_prob} for the intersection area of two circles and after some algebraic manipulations shown in \cite{bkaufman_t1}, we can define the intersection area $A_{INT}$ as
\begin{eqnarray} 
A_{INT}\!\!\!&= \;\Re\Bigg[d_{max}^2\, \cos^{-1}\left[\dfrac{d_{SC}^2 + d_{max}^2 - r^2}{2 d_{max}d_{SC}}\right] +\nonumber
\\
& {}
r^2\, \cos^{-1}\left[\dfrac{d_{SC}^2 - d_{max}^2 + r^2}{2 rd_{SC}}\right] - \nonumber 
\\
&  {}
d_{SC}\sqrt{r^2-\dfrac{(d_{SC}^2-d_{max}^2+r^2)^2}{4d_{SC}^2}}\; \; \Bigg], 
\label{eq:m3_tx_area}
\end{eqnarray}
which gives the area of the source's coverage region that intersects with the cluster.  We note that the formula for the area involves taking the real component of a complex answer.  The notion of a complex area may seem strange, however the formula was derived under the assumption that the two circle's edges intersect each other.  In the scenario where one circle is completely contained within the other, we know that there is a complete overlap of the areas.  The trigonometric functions in the formula give a complex result where the real part of the complex area is exactly the area of the smaller circle contained within the boundary of the larger.  

From these discussions, we are finally able to derive the probability of a D2D link existing between two randomly placed D2Ds in the cluster on a given channel $C_i$.  As will be shown below, the probability of outage $P^{out}_{D|C_{i}}$ is a function of nine random variables.  Thus to derive the final outage expression, nine probability distributions need to be averaged over.  We will first give the final outage expression in terms of $Pr[d_{SD} \leq d_{max}]$ and $A_{INT}$ from above and then decompose the expression into smaller parts.  This well help for both paper organization and mathematical intuition.  

As shown above, $Pr[d_{SD} \leq d_{max}]$ gives the complement of the outage probability and is is expressed in terms of the ratio of $A_{INT}$ to the area of the cluster, $\pi r^2$.   Averaging over all the channel gains and distances found in (\ref{eq:d}) and (\ref{eq:m3_tx_area}) gives (\ref{eq:outage_general})
which is the D2D link outage probability as a function of the random network parameters.  By definition, the Rayleigh fading terms are independent of the distances.  This allows us to write
\setcounter{equation}{12}
\begin{IEEEeqnarray}{rCl}
p(&h&_{MB},h_{MD},h_{SD},d_{SC},d_{SB},d_{MD},d_{MB})=\IEEEnonumber\\
&&p(h_{MB},h_{MD},h_{SD})\; p(d_{SC},d_{SB},d_{MD},d_{MB}),
\end{IEEEeqnarray}
which decomposes the six variable probability distribution function in (\ref{eq:outage_general}) to a product of two smaller distribution functions.  Furthermore, each of the channel gains are characterized by independent and identical exponential distributions and when combined give
\begin{eqnarray}
p(h_{MB},h_{MD},h_{SD}) &=& p(h_{MB})\;p(h_{MD})\;p(h_{SD}) \nonumber
\\
&=&\prod_{i=1}^3 \dfrac{1}{\upsilon}e^{-h_i/\upsilon},
\label{eq:h_marginals}
\end{eqnarray}
and each has mean $E[h_{ij}]=\upsilon$.

We now decompose the joint probability distribution $p(d_{SC},d_{SB},d_{MD},d_{MB})$ into a product of functions to show the various interdependencies the distances have on each other.  We denote the distance between the center of the randomly located cluster of radius $r$ and the base station to be $d_{CB}$.  To satisfy the requirement that the cluster and all the D2Ds within the cluster are located inside the macrocell of radius $R$, we use a shifted uniform distribution, expressed as
\begin{equation} 
p(d_{CB}) = \dfrac{2d_{CB}}{(R-r)^2}, \;\;\; \text{for} \;\;\; 0\leq d_{CB} \leq R-r, 
\label{eq:dcb}
\end{equation}
to statistically characterize $d_{CB}$.  Let the distance from the D2D source to the center of the cluster be $d_{SC}$, and is characterized by the standard uniform distribution
\begin{equation} 
p(d_{SC}) = \dfrac{2d_{SC}}{r^2}, \;\;\; \text{for} \;\;\; 0\leq d_{SC} \leq r. 
\label{eq:dsc}
\end{equation} 
The location of the cluster and D2D source constrain the domain of feasible values for the distance between the source and the base station, denoted as $d_{SB}$.  We can express this conditional dependence on $d_{SC}$ and $d_{CB}$ with the distribution
\begin{equation} 
p(d_{SB}|d_{SC},d_{CB}) = 
\begin{cases}
\dfrac{2d_{SB}}{\pi d_{SC}^2} \cos^{-1}(\psi), 
\\
\;\;\;\;\;\;\;\;\;\;\;\;\;\;\text{for} \;\;\; |d_{SB} - d_{CB}|\leq d_{SC},
\vspace{10pt}
\\
0, \;\;\; \text{elsewhere},
\end{cases}
\label{eq:m3_D_pdf}
\end{equation}
where $\psi$ is defined as
\begin{equation}
\psi = \dfrac{d_{SB}^2+d_{CB}^2-d_{SC}^2}{2d_{SB}d_{CB}},
\label{eq:dsb}
\end{equation}
as shown in \cite{geometry_prob}.

We now characterize the three distances that relate the macro user to the D2D cluster.  Macro users are uniformly distributed inside the cell and the distance from a macro user to the base station, $d_{MB}$, follows the standard uniform distribution
\begin{equation} 
p(d_{MB}) = \dfrac{2d_{MB}}{R^2}, \;\;\; \text{for} \;\;\; 0\leq d_{MB} \leq R. 
\label{eq:dmb}
\end{equation}
The macro user and the center of the cluster are separated by a distance $d_{CM}$ and are both uniformly distributed within the same circle of radius $R$.  A well known geometry result in \cite{geometry_prob} gives the probability distribution 
\begin{equation} 
p(d_{CM}) =
\begin{cases}
\dfrac{2d_{CM}}{R^2}, \;\;\; \text{for} \;\;\; 0\leq d_{CM} \leq r
\vspace{10pt}
\\
\dfrac{d_{CM}\big(2 \theta - \sin(2\theta) \big)}{\pi R^2} + \dfrac{d_{CM}\big(2 \phi - \sin(2\phi) \big)}{\pi (R-r)^2}, 
\\
\;\;\;\;\;\;\;\;\;\;\;\;\;\;\;\;\;\;\;\;\;\;\;\;\;\;\;\;\;\; \text{for} \;\;\; r \leq d_{CM} \leq 2R-r\\
0, \;\;\; \text{elsewhere}
\end{cases}
\label{eq:dcm}
\end{equation}
where $\theta$ and $\phi$ are defined as 
\begin{eqnarray} 
\theta = \cos^{-1}\Bigg[\dfrac{d_{CM}^2+r^2-2Rr}{2d_{CM}(R-r)}\Bigg],
\\
\phi = \cos^{-1}\Bigg[\dfrac{d_{CM}^2-r^2+2Rr}{2d_{CM}R}\Bigg],
\end{eqnarray}
for the distance between two uniformly random points in the same circle.  The location of the macro user with respect to the cluster limits the domain of feasible values for the distance between the macro user and the D2D destination, denoted as $d_{MD}$.  We can express this conditional dependence on $d_{CM}$ with the distribution
\begin{equation} 
p(d_{MD}|d_{CM}) = 
\begin{cases}
\dfrac{2d_{MD}}{\pi r^2} \cos^{-1}\Bigg[ \dfrac{d_{MD}^2+d_{CM}^2-r^2}{2d_{MD}d_{CM}}\Bigg],
\\
\;\;\;\;\;\;\;\;\;\;\;\;\;\;\;\;\;\;\;\;\;\;\;\;\; \text{for} \;\; |d_{MD} - d_{CM}|\leq r
\vspace{10pt}
\\
0, \;\;\; \text{elsewhere}
\end{cases}
\label{eq:dmd}
\end{equation}
as shown in \cite{geometry_prob}.

Using the probability distributions given above, we can write 
\begin{IEEEeqnarray}{rCl}
p(d_{SC},&d&_{SB},d_{MD},d_{MB}) = \IEEEnonumber
\\
&&p(d_{SB},d_{SC},d_{CB})\;p(d_{MB})\;p(d_{MD},d_{CM}),
\label{eq:dist_marginals}
\end{IEEEeqnarray}
where
\begin{IEEEeqnarray}{rCl}
p(d_{SB},d_{SC},d_{CB}) = p(d_{SB}|d_{SC},d_{CB})\;p(d_{SC})\;p(d_{CB}),\;\;\;\;\;\;\;
\label{eq:dist_marginals2}
\end{IEEEeqnarray}
\begin{IEEEeqnarray}{rCl}
p(d_{MD},d_{CM}) = p(d_{MD}|d_{CM})\;p(d_{CM}),
\label{eq:dist_marginals3}
\end{IEEEeqnarray}
follows from Bayes' theorem.  Thus using (\ref{eq:dist_marginals}), (\ref{eq:dist_marginals2}), and (\ref{eq:dist_marginals3}), the joint probability distribution for the distances related to the outage probability $P^{out}_{D|C_{i}}$ can be written as a product of six closed form probability distributions.  Evaluating the probability of outage in (\ref{eq:outage_general}) requires integrating over the nine random variables found in (\ref{eq:h_marginals}) and (\ref{eq:dist_marginals}) and obtaining a closed form expression for the outage probability is intractable.  Therefore the final analytical expressions are numerically approximated and verified through simulation.  
\subsection{Bounds on the Single-hop Probability of Outage}
\begin{figure}[htp]
\center
 \includegraphics[scale=0.83]{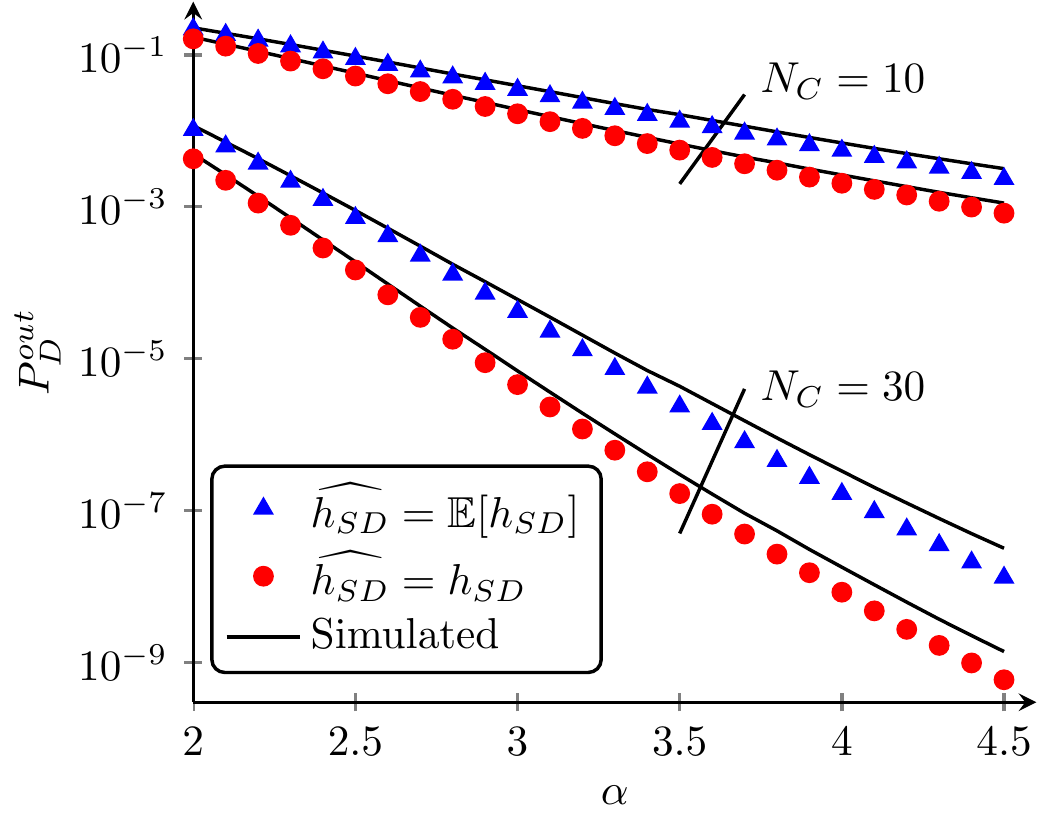}
  \caption[]{Analytical and simulated results for the probability of outage of a single D2D link ($P^{out}_{D}$) on any of $N_C$ channels, each with a single interfering macro user, for a radius ratio of $r/R = 0.25$. Perfect channel gain estimates and statistical estimates in the D2D source's power control are considered. } 
  \label{fig:clustprobmultichan}
\end{figure}

The exact model as shown in Fig.~\ref{fig:analyticalmodel} was simulated for $10^6$ random topologies.  The network parameters used for the evaluation of the analytical expressions and for the simulations are shown in Table~\ref{table:vars}.  In Fig.~\ref{fig:clustprobmultichan}, we plot the probability of outage in (\ref{eq:m2a_multi_channel_prob3}) for a single D2D link sharing any of the $N_C$ channels with an active macro user versus the pathloss exponent $\alpha$.
We use a fixed radius ratio of ${r}/{R} = 0.25$ and vary the number of channels in the network.  We first consider a perfect channel estimate of $\widehat{h_{SD}}=h_{SD}$ for the power control which gives a lower bound on the outage probability.  Perfect knowledge of the channel is difficult to obtain in practice so our protocol uses a statistical estimate of $\widehat{h_{SD}}=\mathbb{E}[h_{SD}]$ in the power control.  We can see that as $N_C$ increases, $P^{out}_{D}$ decreases as D2Ds have more diversity in the resources that they can use.  The increased diversity allows them to choose a channel with lower interference from macro users.  This in turn makes fading effects along the source-destination link, $h_{SD}$, more significant.  This explains the increasing gap away from the lower bound with increasing $N_C$.  It can also be seen that the analytical results match well with the simulation results and thus we will only show the analytical results in Figs. 6-9.  

\begin{figure}[htp]
\center
 \includegraphics[scale=0.83]{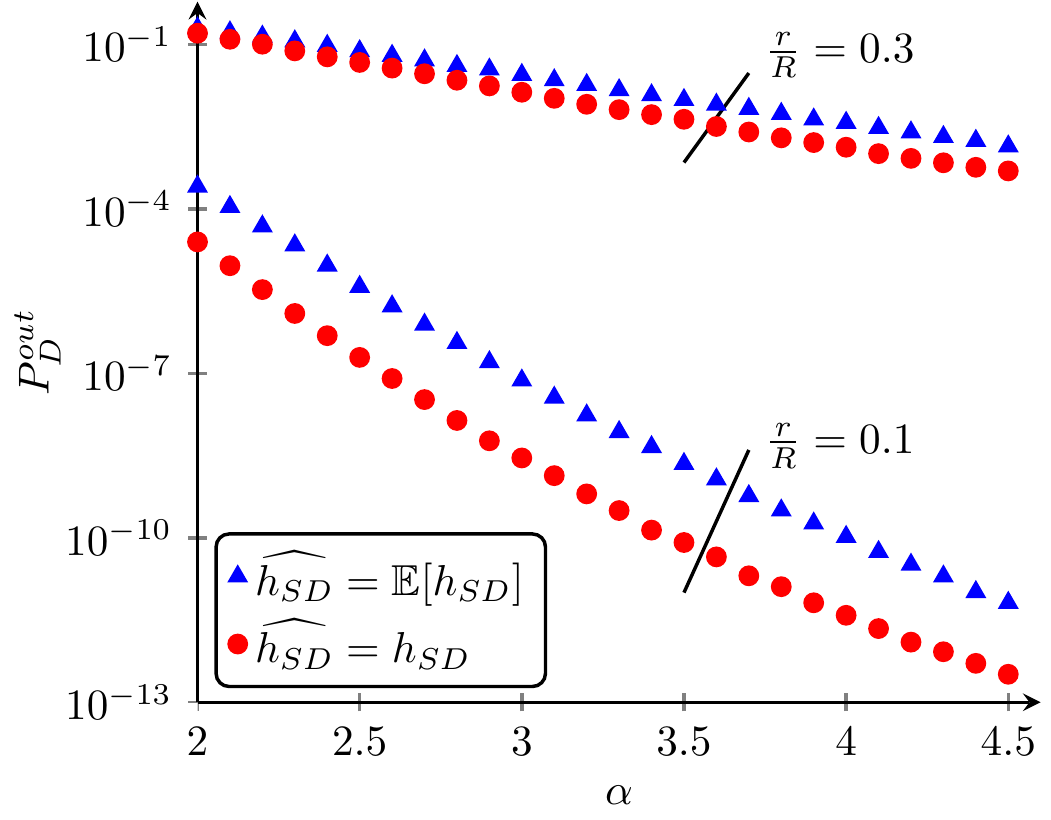}
  \caption[]{Results for the probability of outage of a single D2D link ($P^{out}_{D}$) sharing one of $N_C=15$ channels, each with an active macro user, and for a varying radius ratio $r/R$. Perfect channel gain estimates and statistical estimates in the D2D source's power control are considered. }
  \label{fig:clustprob}
\end{figure}

In Fig.~\ref{fig:clustprob}, we plot the same outage probability in (\ref{eq:m2a_multi_channel_prob3}) with a fixed $N_C = 15$ as the radius ratio $r/R$ varies.
As the size of the cluster decreases relative to the size of the macrocell, the D2D outage probability decreases.  A smaller cluster radius reduces the maximum distance allowed between the D2D source and destination thus improving link quality.  Furthermore, as the source and destination move closer to each other on average, the fading effects in the link become more dominant than the pathloss.  This explains the increasing gap from the statistical estimate from the lower bound as $r/R$ decreases.  We note that in both Fig.~\ref{fig:clustprobmultichan} and Fig.~\ref{fig:clustprob}, the outage probability decreases as the pathloss increases.  With increasing $\alpha$, D2Ds can transmit at a higher power due to the reduction in their interference at the base station.  They will also receive less interference from an interfering macro user.  
\begin{figure}[htp]
\center
 \includegraphics[scale=0.83]{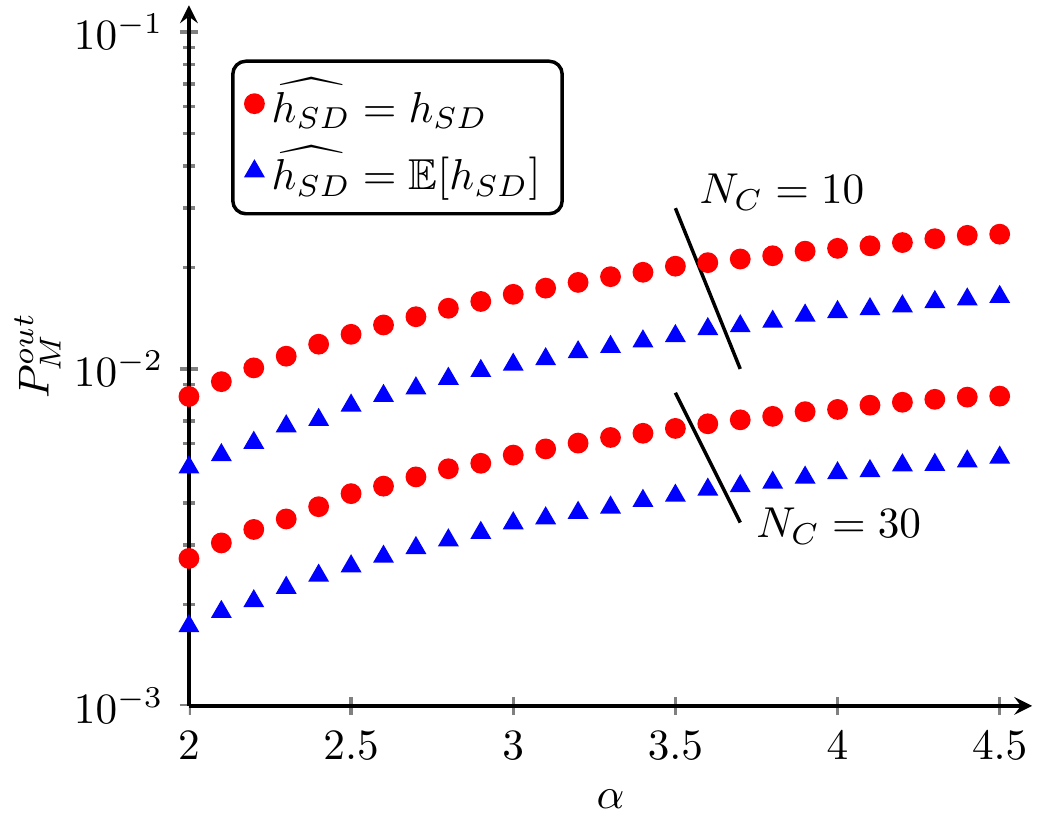}
  \caption[]{Results for the probability of outage of $N_C$ macro users ($P^{out}_{M}$) sharing their channel with a single D2D link for a radius ratio of $r/R = 0.25$. Perfect channel gain estimates and statistical estimates in the D2D source's power control are considered. }
  \label{fig:MUprobmultichan}
\end{figure}

\begin{figure}[htp]
\center
 \includegraphics[scale=0.83]{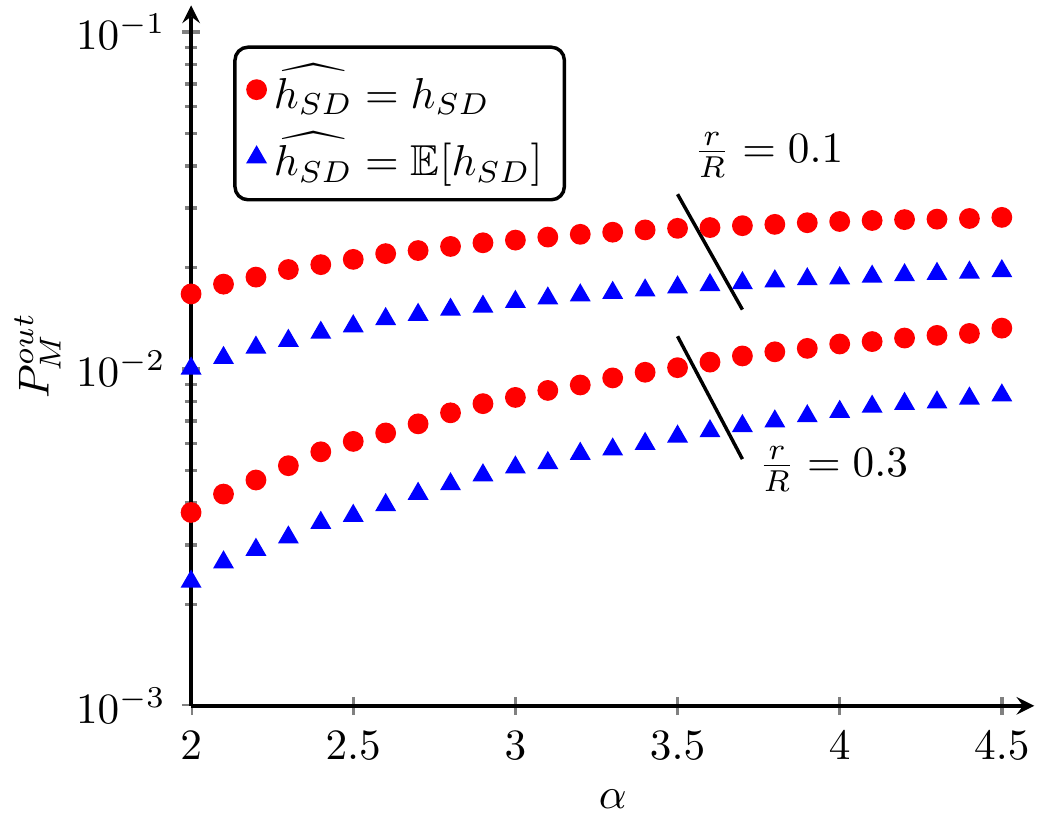}
  \caption[]{Results for the macro user probability of outage ($P^{out}_{M}$) for $N_C = 15$ macro users sharing their channel with a single D2D link for a varying radius ratio $r/R$. Perfect channel gain estimates and statistical estimates in the D2D source's power control are considered. }
  \label{fig:MUclustprob}
\end{figure}

\begin{figure}[htp]
\center
 \includegraphics[scale=.85]{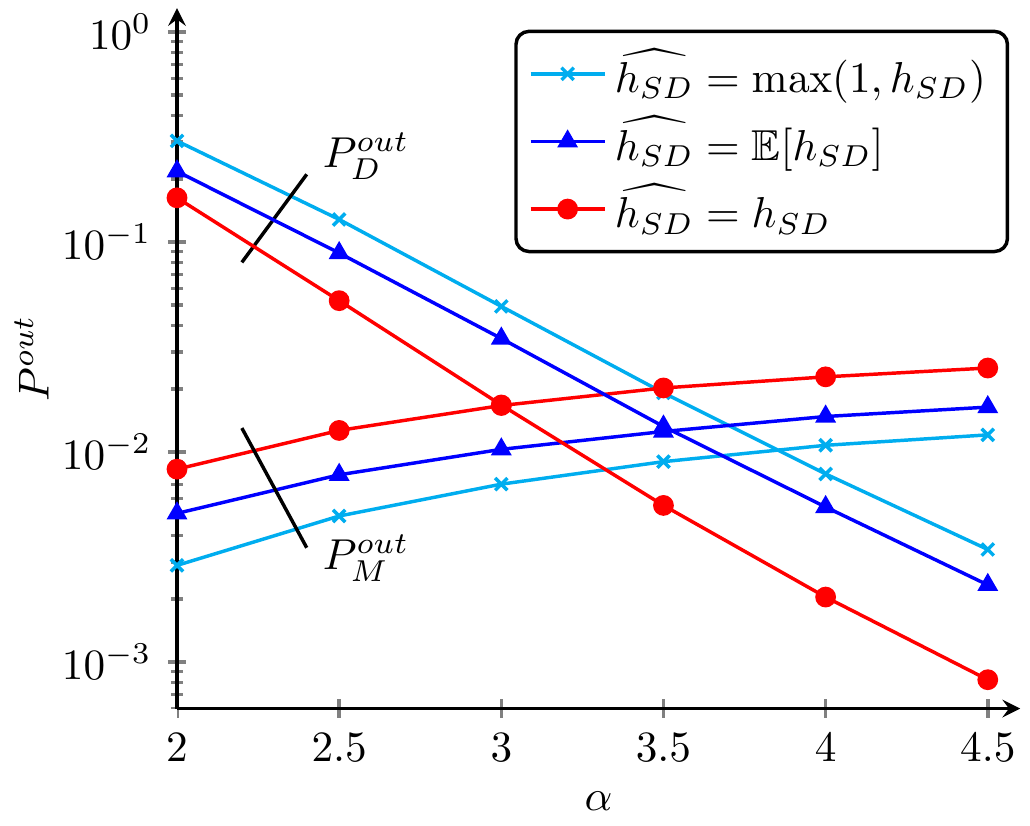}
  \caption[]{Results for the probability of outage for D2D users ($P_{D}^{out}$) and macro users ($P_{M}^{out}$) for $N_C=10$ and $r/R=0.25$.  Power control using a statistical estimate is upper and lower bounded by two different perfect channel inversion techniques. }
  \label{fig:truncatedPC}
\end{figure}

In Fig.~\ref{fig:MUprobmultichan} and Fig.~\ref{fig:MUclustprob}, we plot the analytical probability of outage for a macro user, $P_M^{out}$, who is sharing its channel with a D2D link.  The calculations to derive $P_M^{out}$ are done in the same manner as that for a D2D link, however they are much easier due to the fixed location of the base station.  Due to these reasons, as well as space constraints, we omit the exact expressions for $P_M^{out}$.  We include these results for completeness as it is important to quantize the affect D2Ds have on the cellular network.  We see that as either $N_C$ or $r/R$ varies, most values of $P_M^{out}$ are less than $10^{-2}$ and start to level off for large $\alpha$.  
In order for the D2D outage probability to decrease with $\alpha$, it is expected for the macro user outage probability to increase.  This is clearly shown in Fig.~\ref{fig:truncatedPC} where we plot the analytical outage probability for macro users and D2D users together.  We see that D2Ds get about a factor of 100 improvement while the macro user loss is significantly less than a factor of 10.  
We now also plot the outage using the truncated channel inversion in the power control, $\widehat{h_{SD}}=\max(1,h_{SD})$, which was described above.  This is an analytical upper bound on the D2D outage where it shows the best performance of a D2D link while guaranteeing that the power control over the D2D source-destination link does not cause outage at the base station.  In this scenario, the outage of the macro user link comes from the estimate of the channel gain between the D2D source and the base station.  We also note that as the different channel gain estimates cause the D2D outage to increase, there is a corresponding decrease in the macro user outage as there is always a tradeoff in performance as the two different classes of users try to share the spectrum.  

\subsection{Multi-hop Simulation Results}
\label{sec:multihop_simulation}

The performance of the network can be further quantified by looking at the power savings when D2D communication is used instead of communicating using the standard cellular mode.  To quantify this, we consider the cellular mode power to be the sum of the powers for the D2D source to reach the base station, $P_{T_{SB}}$, and for the base station to reach the D2D destination, $P_{T_{BD}}$.  We calculate the D2D mode power for a route of length $N_{Hops}$ as the sum of the D2D's transmit power used in the route where the $n$'th D2D transmits with power $P_{T_{D_n}}^*$.  
\begin{figure}[htp]
\center
  \includegraphics[scale=.85]{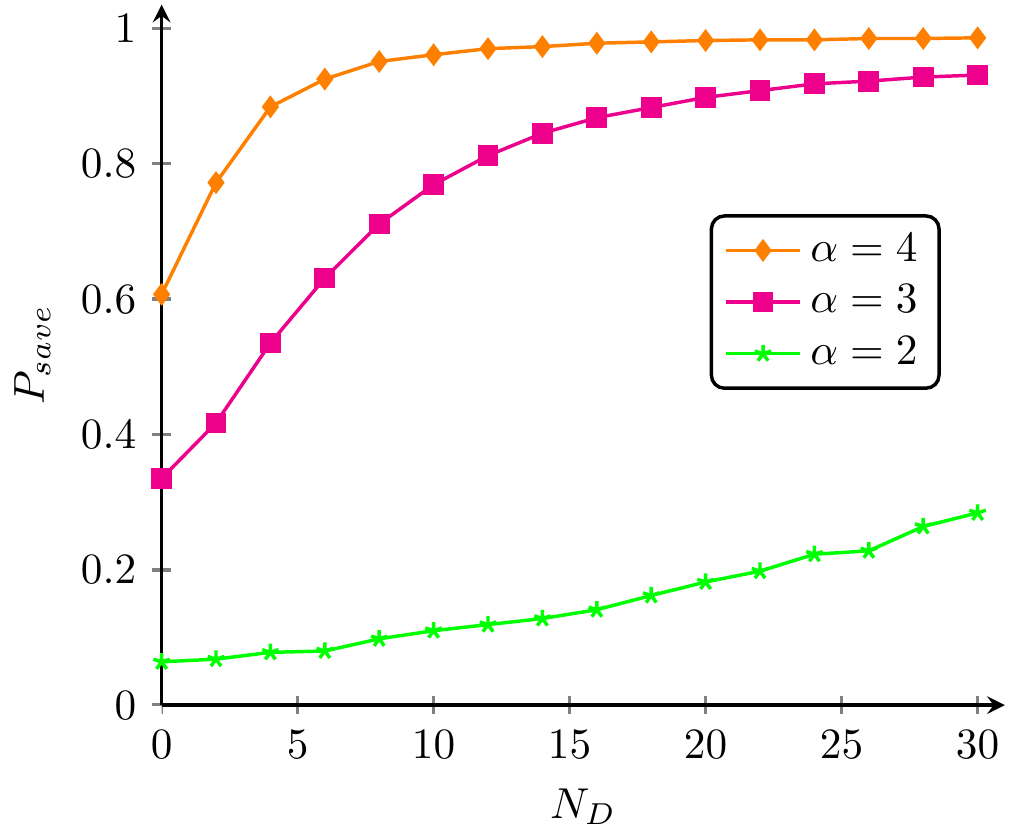}
  \caption[]{Significant power savings ($P_{save}$) can be seen when using a multi-hop D2D route rather than the cellular mode in high pathloss environments.  Low pathloss environments result in meager savings as there is often a strong channel with the base station.} 
  \label{fig:d2dpower}
\end{figure}   
Using these powers, we calculate the power savings as
\begin{equation}
P_{save} = \dfrac{P_{T_{SB}}+P_{T_{BD}}-\sum\limits_{n=1}^{N_{Hops}} P_{T_{D_n}}^*}{P_{T_{SB}}+P_{T_{BD}}},
\end{equation}
and plot the results in Fig.~\ref{fig:d2dpower}.  We note that $N_D=0$ corresponds to a single-hop route between the source and destination.  All values of $\alpha$ achieve significant savings but for $\alpha \geq 3$ in particular, savings of almost 90\% can be achieved for moderate $N_D$.  These power savings come from the fact that D2Ds can communicate over shorter distances more efficiently than longer links with the base station.  When $\alpha$ is larger, D2Ds become more isolated from the base station making shorter distance hops more efficient.  

\begin{figure}[htp]
\center
 \includegraphics[scale=.85]{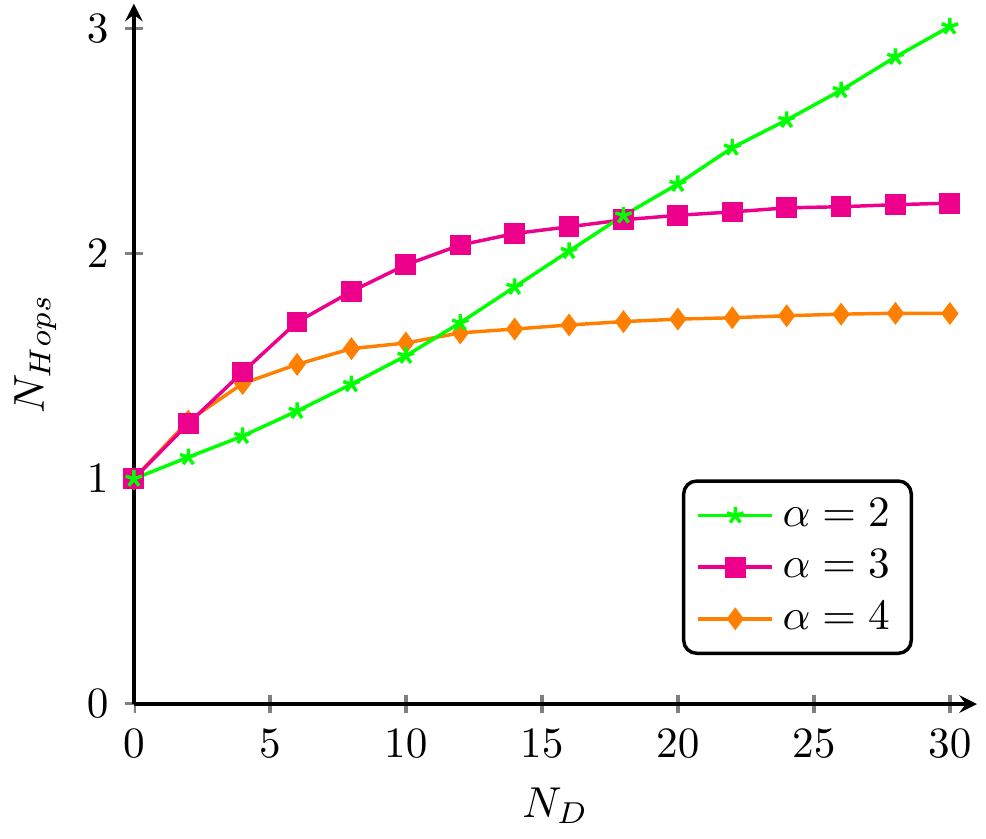}
  \caption[labelInTOC]{The average number of hops ($N_{Hops}$) for a multi-hop D2D route connecting a random source-destination pair.  In small pathloss environments, D2D users must lower their transmit power to limit the interference to the base station, resulting in routes with many hops.  }
  \label{fig:hops}
\end{figure}

We mentioned above that not all of the $N_D$ relays in the cluster participate in the multi-hop route.  We now describe the types of routes that are being formed between the D2D source and destination.  We do this by looking at the average number of hops, $N_{Hops}$, per found route versus the number of D2D relays, as shown in Fig.~\ref{fig:hops}.  
We see that even as $N_D$ increases to large values and for $\alpha \geq 3$, the route length quickly saturates to a low number of hops and tends to be two hops or less.  In the high interference scenario for $\alpha = 2$, routes will quite often span higher number of hops in order to establish a route.  This result shows us that the distances the source and destination are trying to span are not significantly larger than the single-hop distance and usually one to two relays can suffice in establishing a two-way D2D route.  As a further note, we can comment on the actual physical distance that each multi-hop route is able to span.  Recall that we are considering a D2D cluster of radius 500m which corresponds to a maximum separation of 1000m between the D2D source and destination.  With the average number of hops being near two, each hop is capable of spanning up to a few hundred meters.

\section{Conclusions}
\label{sec:conclusion}
In this paper we have presented an opportunistic communication scheme in which an ad-hoc Device-to-Device network can simultaneously communicate on the same set of frequency resources as a fully loaded cellular radio network.  We develop a practical protocol for D2Ds to use this scheme in a distributed manner and with no coordination from the base station.  The D2D users first step is to control their powers to a level which causes minimal interference to the base station.   Then using the calculated power, the second step is to employ a discovery protocol to establish a route connecting them to their intended destination.  

Results show that including network information in the discovery packet significantly lowers the route discovery's failure probability and reduces the number of transmissions necessary to discover a route to the destination.  Given that a route is found, the probability of outage for a D2D link is derived and lower bounded using perfect channel inversion in the power control.  Using a practical statistical estimate in the power control, our protocol shows performance near to the lower bound.  The spectrum is fully utilized by the macro user network so there is a clear tradeoff in the performance of the two classes of users.  However, large improvements in the D2D performance come at a cost of only a small loss in macro user performance.  Furthermore, simulation results show that significant power savings can be gained using D2D routes rather than connecting to the cellular base station.  

To further improve the work, more coordination between the base station and D2D users could be considered.  Currently there is no specific signaling between them which enables the D2D users to be transparent to the cellular network, but makes it difficult for multiple D2D clusters to communicate simultaneously and not overwhelm the base station with too much interference.  An additional extension would be to consider a more dynamic channel estimation for the power control.  Training sequences could be included in discovery packets allowing D2D users to have more accuracy in their power control to guarantee minimal interference to the base station.  We would expect to see even more gains than those presented here if these extensions are considered.  

\ifCLASSOPTIONcaptionsoff
  \newpage
\fi



%
%
%

\bibliographystyle{IEEEtran} 
\bibliography{journal}

\end{document}